\documentclass[useAMS,twocolumn]{mn2e}
\usepackage{amssymb}
\usepackage{graphicx}

\def\be{\begin{equation}}
\def\ee{\end{equation}}

\catcode`\@=11 
\def\@versim#1#2{\vcenter{\offinterlineskip
        \ialign{$\m@th#1\hfil##\hfil$\crcr#2\crcr\sim\crcr } }}
\def\mpy{M_{\sun} \ {\rm yr^{-1}}}

\usepackage[dvips]{color}

\begin{document}

\title[Convective instability of hot radiative accretion flows]
{On the convective instability of hot radiative accretion flows}
\author[Yuan \& Bu]
{Feng Yuan$^{1}$\thanks{E-mail:fyuan@shao.ac.cn } and De-Fu Bu$^{1,2}$\thanks{E-mail:dfbu@shao.ac.cn}\\
$^{1}$Key Laboratory for Research in Galaxies and Cosmology,
Shanghai Astronomical Observatory,\\ Chinese Academy of
Sciences, 80 Nandan Road, Shanghai 200030, China\\
$^{2}$Graduate School of the Chinese Academy of Sciences, Beijing
100039, China\\}

\maketitle

\date{Accepted . Received ; in original form}

\markboth{}{}

\begin{abstract}
How many fraction of gas available at the outer boundary can finally
fall onto the black hole is an important question. It determines the
observational appearance of accretion flows, and is also related
with the evolution of black hole mass and spin. Previous
two-dimensional hydrodynamical simulations of hot accretion flows
find that the flow is convectively unstable because of its inward
increase of entropy. As a result, the mass accretion rate decreases
inward, i.e., only a small fraction of accretion gas can fall onto
the black hole, while the rest circulates in the convective eddies
or lost in convective outflows. Radiation is usually neglected in
these simulations. In many cases, however, radiative cooling is
important. In the regime of the luminous hot accretion flow (LHAF),
radiative cooling is even stronger than the viscous dissipation. In
the one dimensional case, this implies that the inward increase of
entropy will become slower or the entropy even decreases inward in
the case of an LHAF. We therefore expect that convective instability
becomes weaker or completely disappears when radiative cooling is
important. To examine the validity of this expectation, in this
paper we perform two-dimensional hydrodynamical simulations of hot
accretion flows with strong radiative cooling. We find that compared
to the case of negligible radiation, convection only becomes
slightly weaker. Even an LHAF is still strongly convectively
unstable, its radial profile of accretion rate correspondingly
changes little. We find the reason is that the entropy still
increases inward in the two-dimensional case.
\end{abstract}


\section{INTRODUCTION}

Hot accretion flows such as advection-dominated accretion flows
(ADAFs) are interesting because they are likely operating in
low-luminosity active galactic nuclei (AGNs) and hard and quiescent
states of black hole X-ray binaries (see Narayan 2005; Yuan 2007;
Narayan \& McClintock 2008 for recent reviews). ADAFs are originally
proposed and studied by vertically-integrated one-dimensional method
since the global two-dimensional solution is technically too
difficult to obtain (e.g., Narayan \& Yi 1994; 1995; Abramowicz et
al. 1995). While this approach has discovered the main properties of
ADAFs, some important multi-dimensional effects such as convection
and outflow as we will focus in this paper are neglected and await
the multi-dimensional numerical simulations.

Perhaps the most important finding of multi-dimensional
hydrodynamical simulations is that the flows are highly convectively
unstable (Igumenshchev \& Abramowicz 1999, hereafter IA99; Stone,
Pringle \& Begelman 1999, hereafter SPB99; Igumenshchev \&
Abramowicz 2000), in consistent with what suggested by
one-dimensional self-similar solution of ADAFs (Narayan \& Yi 1994;
1995). The physical reason is that the entropy of the accretion flow
increases inward. Because of the convective instability, most of the
gas available at the outer boundary can't fall onto the horizon of
the central black hole, but circulates in convective eddies or lost
in convective outflows. As a result, the mass accretion rate keeps
decreasing inward.

The radial profile of accretion rate has important observational
implications. The radiative appearance of black hole obviously
depends on it. This is crucial for us to explain some observations.
The first example is the supermassive black hole in the Galactic
center, Sgr A*. {\em Chandra} observations combined with Bondi
accretion theory present a robust estimation to the value of
accretion rate at the outer boundary of the accretion flow, the
Bondi radius, which is about $10^{-5}\mpy$ (Baganoff et al. 2003).
On the other hand, the detected high linear polarization at radio
waveband requires a mass inflow rate of only $10^{-7}$-$10^{-9}\mpy$
at the innermost region of the ADAF (Marrone et al. 2007). So most
of the gas can't fall onto the black hole. The second example is the
transition from hard to soft states of black hole X-ray binaries.
The model of the hard state is an inner hot accretion flow plus an
outer truncated thin disk outside of the truncation radius $R_{\rm
tr}$, while the soft state is described by a standard thin disk. One
well-known observational result is that the luminosity changes
little during the transition (e.g., Zdziarski et al. 2004). This
requires that the accretion rate of the inner hot accretion flow
should not decrease significantly inward from $R_{\rm tr}$, because
otherwise we would expect that the luminosity had increased
significantly after the transition due to the higher accretion rate
of the thin disk. In addition to the effects on the emitted
spectrum, the exact profile of accretion rate is also important to
the study of evolution of black hole mass and spin at least in the
phase of low-luminosity AGNs (LLAGNs), in which a hot accretion flow
is believed to be working (Ho 2008).

In all numerical simulations mentioned above, however, radiation is
neglected. This is a good approximation only when the mass accretion
rate is very low so radiative cooling is unimportant. In reality,
the accretion rate is often high enough thus radiative cooling can't
be neglected. Moreover, when the accretion rate is high enough, such
as in the luminous hard state and LLAGNs, the flow will enter into
the regime of luminous hot accretion flow (LHAF; Yuan 2001; 2003).
In this case, the radiative cooling rate is higher than the viscous
dissipation rate. In one-dimensional case, this implies that the
entropy of the gas {\em decreases} inwardly (see Yuan 2001 or \S3.2
in this paper for details). Therefore, different from an ADAF, an
LHAF is predicted to be convectively stable (Yuan 2001).

In the present work, we simulate the two-dimensional hydrodynamical
accretion flow with radiation. We do this by adding bremsstrahlung
radiation in the energy equation. Our purpose is to examine the
effects of radiation on the dynamics of hot accretion flow,
especially on the convective instability. Surprisingly, our results
indicate that the accretion flow is still convectively unstable,
even when the flow is in the regime of an LHAF. In Section 2, we
describe our numerical method. Results are described in Section 3.
We summarize and discuss our results in Section 4.

\section{METHOD}

\subsection{The equations of motion}
The hydrodynamical equations describing accretion including
bremsstrahlung radiation are:
\begin{equation}
\frac{d\rho}{dt}+\rho\nabla\cdot \mathbf{v}=0,\label{cont}
\end{equation}
\begin{equation}
\rho\frac{d\mathbf{v}}{dt}=-\nabla p-\rho\nabla
\psi+\nabla\cdot\mathbf{T}, \label{rmon}
\end{equation}
\begin{equation}
\rho\frac{d(e/\rho)}{dt}=-p\nabla\cdot\mathbf{v}+\mathbf{T}^2/\mu-Q^{-}_{\rm
rad}, \label{rmon}
\end{equation}
where $\rho$, $p$, $\psi$, ${\bf v}$, $e$ and ${\bf T}$ are density,
pressure, gravitational potential, velocity, internal energy and
anomalous stress tensor, respectively. The
$d/dt(\equiv\partial/\partial t + \mathbf v \cdot\nabla)$ denotes
the Lagrangian time derivative. We adopt an equation of state of
ideal gas $p=(\gamma -1)e$, and consider models with $\gamma =5/3$.
We assume bremsstrahlung cooling $Q^{-}_{\rm
rad}=6.2\times10^{20}\rho^2T^{1/2} {\rm erg}\cdot {\rm s}^{-1}\cdot
{\rm cm}^{-3}$, with $T$ is the temperature of the accretion flow.

We use the stress tensor $\bf T$ to mimic the shear stress which is
in reality magnetic stress associated with MHD turbulence driven by
the magneto-rotational instability (MRI; Balbus \& Hawley 1998).
Following SPB99, in most cases, we assume that the only non-zero
components of $\bf T$ are the azimuthal ones:
\begin{equation}
  T_{r\phi} = \mu r \frac{\partial}{\partial r}
    \left( \frac{v_{\phi}}{r} \right),
\end{equation}
\begin{equation}
  T_{\theta\phi} = \frac{\mu \sin \theta}{r} \frac{\partial}{\partial
  \theta} \left( \frac{v_{\phi}}{\sin \theta} \right) .
\end{equation}
This is because the MRI is driven only by the shear associated with
orbital dynamics. But in order to compare with previous works we
also use the full components in the case of large $\alpha$ (\S3.3).
Following SPB99, in most calculations we adopt the coefficient of
shear viscosity $\mu=\nu\rho$ and the magnitude of the kinematic
coefficient $\nu=0.1\rho/\rho_{\rm max}$. Here $\rho_{\rm max}$ is
the maximum density of the initial torus adopted in our simulation
(see \S2.2 below).

We use pseudo-Newtonian potential to mimic the general relativistic
effects, $\psi=-GM/(r-r_s)$, where $G$ is the gravitational
constant, $M$ is the mass of the black hole, and $r_s=2GM/c^2$ is
the Schwarzschild radius. We neglect the self-gravity of the disk.

\subsection{Initial conditions}

Following SPB99, the initial state of our simulations is an
equilibrium torus with constant specific angular momentum. The
equilibrium structure of the torus is given by (Papaloizou \&
Pringle 1984)
\begin{equation}
 \frac{p}{\rho} = \frac{(\gamma-1)GM}{\gamma R_{0}} \left[ \frac{R_{0}}{r} - \frac{1}{2}
\left( \frac{R_{0}}{r \sin \theta} \right)^{2} - \frac{1}{2d} \right] .
\end{equation}
Here $R_{0}$ is the radius of the center (density maximum) of the
torus, and $d$ is the distortion of the torus. As used in SPB99, our
torus is embedded in a low density medium. The density and the
pressure of the medium are $\rho_m$ and $p_m=\rho_m/r$,
respectively.

The units adopted in our calculations are listed in table 1. Note
that we introduce a parameter $n$ in the unit of density. By setting
it to be different values, we can adjust the density of the initial
torus and thus the accretion rate. In this paper, we calculate three
values of $n$, namely $1.2, 0.012$ and $0.00012$. They correspond to
Model A, B, and C, respectively.

\begin{table}
\footnotesize
\begin{center}
\caption{Units adopted in the calculation}
\begin{tabular}{ccc} \\ \hline
Physical quantity
& Symbol
& Numerical unit \\ \hline

Length & $r_g$  & $3\times 10^6 (M/10M_{\odot}) ~{\rm cm}$ \\
Velocity & $c$  & $2.9979\times 10^{10} {\rm cm \cdot s^{-1}}$ \\
Time & $t_0$    & $5\times10^{-5} (M/10M_{\odot})~{\rm sec} $ \\
Density & $\rho_0$ & $2 \times 10^{-7} n (M/10M_{\odot})^{-1} {\rm g \cdot cm^{-3}}$ \\
Temperature & $T_0$ & $1.1\times10^{13} K$ \\
Luminosity & $L_{Edd}$ & $1.25 \times 10^{39} (M/10M_{\odot})~ {\rm erg \cdot s^{-1}}$\\
\hline
\end{tabular}\\
\end{center}
\end{table}

The radius of the maximum density is set as $R_0=90 r_s$, the
specific angular momentum of the initial tours equals to the
Keplerian angular momentum at $R_0$, the maximum density of the
tours $\rho_{max}=0.29$ and the density of the medium
$\rho_m=10^{-4}$.

\begin{figure*}
\begin{center}
\includegraphics[width=0.9\textwidth]{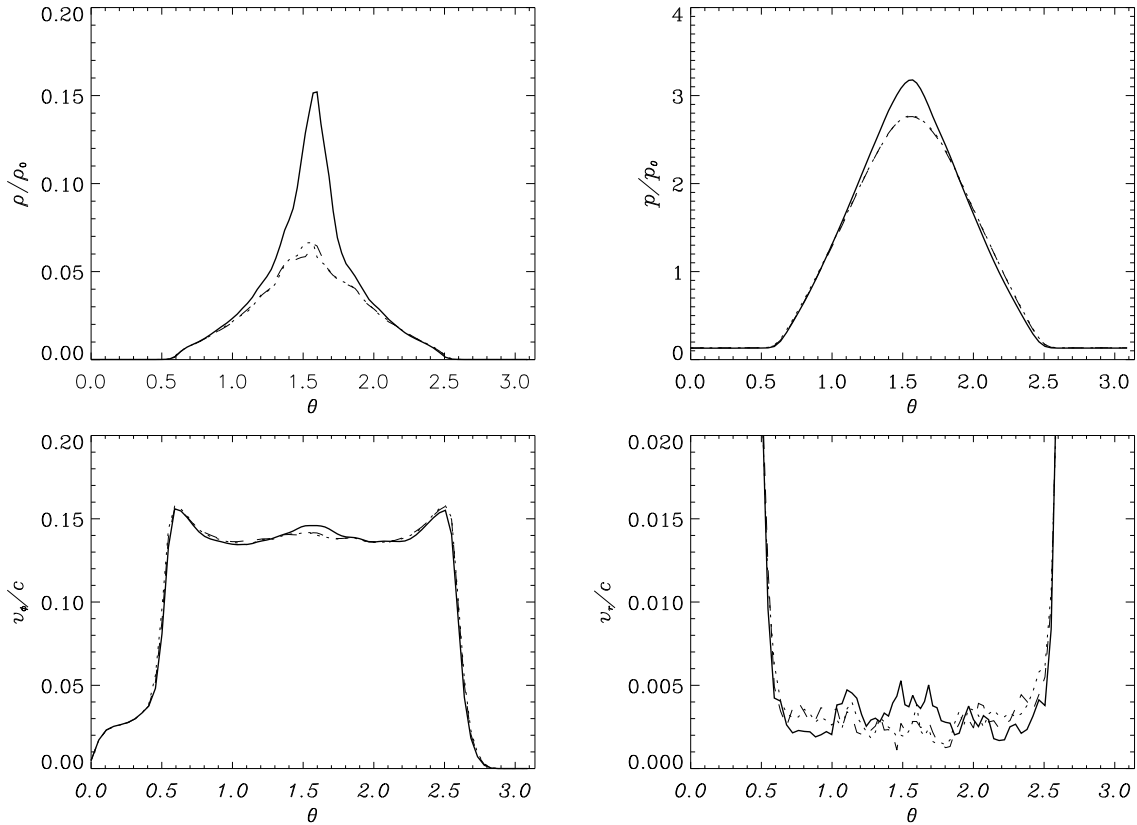}
\end{center}
\caption{Angular profiles of a variety of time-averaged variables at
$r=20r_s$. The solid, dotted and dashed lines correspond to Model A,
B, and C, respectively.}
\end{figure*}

\subsection{Numerical methods}

We use ZEUS-MP code (Hayes et al. 2006) in spherical geometry to
solve the above equations. As a test, we have successfully
reproduced all the results of SPB99. For our calculations, we set
our computation grid extending from an inner boundary at $r=1.2r_s$
to $300r_s$. McKinney \& Gammie (2002) show that the inner boundary
must be smaller than the sonic point of the accretion flow, which is
$\sim 1.5 r_s$, otherwise the location of the inner boundary would
affect the simulation results. We have also tried larger spacial
range and find that the range adopted above does not affect our
results, especially when the effects of radiative cooling are
concerned, which is our main focus of the present work. The number
of grids is \textbf{$(N_r,N_\theta) = (180,100)$}. We adopt
non-uniform grid in the radial direction $(\bigtriangleup r)_{i+1} /
(\bigtriangleup r)_{i} = 1.0183$. Similarly, we adopt non-uniform
angular zones with $(\bigtriangleup \theta)_{j+1} / (\bigtriangleup
\theta)_{j} = 0.9826$ for $0 \leq \theta \leq \pi/2$  and
$(\bigtriangleup \theta)_{j+1} / (\bigtriangleup \theta)_{j} =
1.0177$ for $\pi/2 \leq \theta \leq \pi$. Outflow boundary
conditions are adopted at both the inner and outer radial
boundaries. In the angular direction, the boundary conditions are
set by symmetry at the poles.

\section{RESULTS}

\begin{figure*}
\begin{center}
\includegraphics[width=0.7\textwidth]{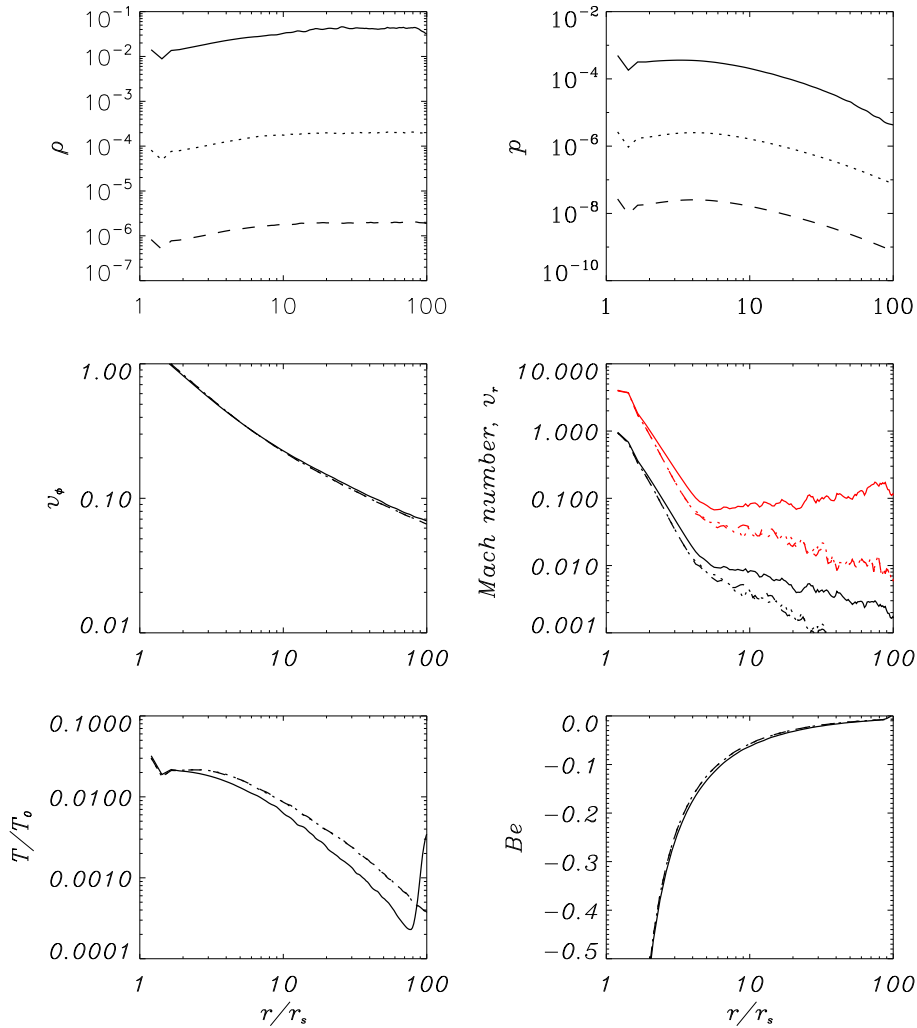}
\end{center}
\caption{Radial scaling of some time-averaged quantities. The
solution is averaged over angle between $\theta=84^{\circ}$ to
$\theta=96^{\circ}$ and over 14 and 16 orbits. In every panel, the
solid, dotted and dashed lines correspond to Model A, B, and C,
respectively. The red and black lines in the middle-right panel
correspond to the Mach number and radial velocity, respectively.}
\end{figure*}

\subsection{The radial and angular structure}

In this paper, time is measured in unit of orbital time at
$r=90r_s$. We switched on the radiative cooling term after $13$
orbital time when the disk becomes quasi-steady. When radiative
cooling is included, we find that it is quicker for the accretion
flow to reach a steady state. This is because when the radiation is
included, the density becomes more concentrated toward the
equatorial plane, as shown by Figure 1. Since the ratio of viscous
stress and other forces is proportional to $\rho/\rho_{\rm max}$,
this results in a quicker establishment of a steady state. The
reason of the concentration of the accretion flow is that the
temperature decreases after the radiation cooling is included thus
the accretion flow condenses.

Figure 1 shows the angular structure of the time-averaged variables
at $r=20r_s$. The solid, dotted and dashed lines correspond to Model
A, B, and C, respectively. The density and gas pressure have been
normalized in these figures. We can see from the figure that the
dotted and dashed lines are almost superposed together, which
implies that the radiative cooling is negligible in these two
models. With increasing mass accretion rates, radiation becomes more
and more important. As a result, the temperature decreases, so the
density profile becomes more concentrated to the equatorial plane,
i.e., the accretion flow becomes thinner. The concentration of
density to the equatorial plane results in a larger radial velocity.
But the effects of radiation on the angular velocities are very
small. This is because the radiation pressure is negligible compared
to the gas pressure.

Figure 2 shows the radial structure of the time-averaged flow near
the equatorial plane. The solution is averaged over angle between
$\theta=84^o$ to $\theta=96^o$. In every panel, the solid, dotted
and dashed lines correspond to Model A, B, and C, respectively. The
red and black lines in the middle-right panel represent the Mach
number and the radial velocity, respectively. In all of our models
the flows have become supersonic before they reach the inner
boundary. This ensures that the location of inner boundary does not
affect our results (McKinney \& Gammie 2002). The density, gas
pressure, rotation velocity and temperature in each model can be
described by a radial power law, with $\rho \propto r^{0}$, $p
\propto r^{-1}$, $v_{\phi} \propto r^{-1/2}$ and $ T \propto r^{-1}
$. Especially, the Bernoulli parameter $Be$($Be\equiv v^2/2+\gamma
P/(\gamma-1)\rho-GM/(r-r_s) $) for the three models are all below
zero.  The negative $Be$ is different from the one-dimensional or
two-dimensional self-similar results (Narayan \& Yi 1994, 1995; see
also Blandford \& Begelman 1999) where the Bernoulli parameter along
the equatorial plane is always positive when radiative cooling is
weak. In our calculations, we set the kinetic viscosity coefficient
$\nu \propto \rho$ which is different from the usual ``$\alpha$''
description adopted in previous works (Narayan \& Yi 1994; 1995)
which corresponds to $\nu \propto r^{1/2}$. But the discrepancy of
the value of $Be$ is not because of the difference of the viscosity
description. SPB99 show that the value of $Be$ is even more negative
when $\nu \propto r^{1/2}$. The discrepancy is because when the
accretion flow is convectively unstable, the scaling law adopted in
the above self-similar solution does not apply any more (Narayan,
Igumenshchev \& Abramowicz 2000; Quataert \& Gruzinov 2000).

The negative nature of $Be$ along the equatorial plane implies that
very few of the accretion flow will escape and flow out to infinity
(IA99; SPB99). To further investigate the strength of outflow, we
show the angular structure of Bernoulli parameter $Be$ at 40 $r_s$
in Figure 3. The angular profile of $Be$ at other radii is similar
to the profile at $40 r_s$. We can see that away from the equatorial
plane, its value becomes larger, consistent with the prediction of
Narayan \& Yi (1995). It becomes positive close to the polar region.
This is because the viscous stress transfer energy from the equator
to the pole. Since the density in that region is very low, we can
expect that the unbound outflow will be extremely weak. Our
simulations show that the mass flux of unbound outflow is only $\sim
1\%$ of the inflow flux. To produce strong unbound outflow, other
mechanism such as large-scale magnetic field is required. Or, we
should properly take into account the radiative transfer, as we will
discuss in \S4.

\begin{figure}
\begin{center}
\includegraphics[width=0.5\textwidth]{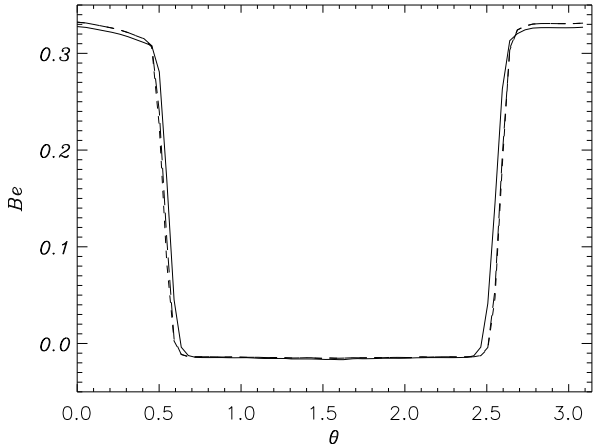}
\end{center}
\caption{The distribution of Bernoulli parameter along the theta
direction at 40 $r_s$. The solid, dotted, and dashed lines
correspond to Model A, B, and C, respectively.}
\end{figure}

Now let's analyze the energetics of the three models. We define the
advection factor of hot accretion flow as the ratio of the energy
advection rate to the viscous dissipation rate:
\begin{equation}
f\equiv \frac{Q_{\rm adv}}{Q_{\rm vis}}=1-\frac{Q_{\rm
rad}^-}{Q_{\rm vis}}.
\end{equation}
Here $Q_{\rm adv}\equiv
\rho\frac{d(e/\rho)}{dt}+p\nabla\cdot\mathbf{v}\equiv \rho T dS/dt$
($S$ is the entropy) is the so-called advection term and $Q_{\rm
vis}\equiv\mathbf{T}^2/\mu$ is the viscous dissipation rate (ref.
eq. 3). Figure 4 shows the time-averaged advection factor near the
equatorial plane of the three models. The result is similar if we
move away from the equatorial plane. We can see from the figure that
$f$ is almost equal to 1 for Models B and C, i.e., they are fully
advection-dominated. This is because the radiative cooling is
negligible compared to viscous dissipation. With the increase of
accretion rates, radiation becomes more and more important. The rate
of viscous heating $$ \mathbf T^2/\mu \propto \mu r^{-3}\propto \rho
r^{-3},$$ and the rate of radiative cooling is
$$Q^-_{\rm rad}\propto \rho^2 T^{1/2}\propto \rho^2 r^{-3/4}.$$ Thus
there exists a critical accretion rate $\dot{M}_{\rm crit}$,
determined by $f =0$, and we have $\dot{M}_{\rm crit}\propto
r^{-9/4}$. Beyond this critical rate, the ADAF solution does not
exist since the radiative cooling is stronger than viscous heating
thus $f<0$ (Abramowicz et al. 1995; Narayan, Mahadevan \& Quataert
1998). This is the case for Model A, where we have $f<0$ for
$r>30r_s$. But note that the accretion flow still remains hot in
this case, as shown in Figure 2. This is because the sum of the
viscous dissipation and compression work is still larger than the
radiative cooling. In this case, the gradient of entropy of
accretion flow, i.e., the left-hand side of the energy equation, is
negative, so advection plays a heating role. This type of accretion
flow is different from ADAFs; instead, it is called the luminous hot
accretion flow (LHAF; Yuan 2001), which is an extension of ADAFs to
higher accretion rates.

\begin{figure}
\begin{center}
\includegraphics[width=0.5\textwidth]{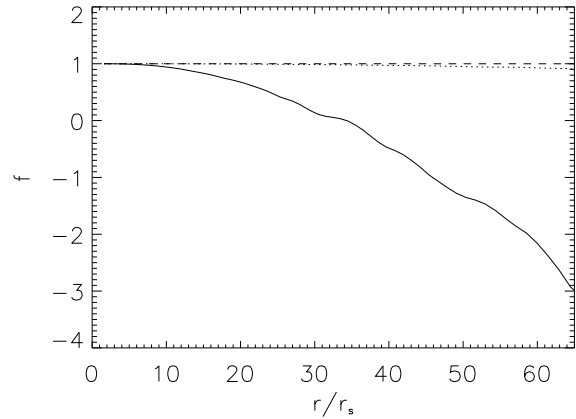}
\end{center}
\caption{Time-averaged advection factor $f$ (ref. eq. 7) near the
equatorial plane. The solution is averaged over angle between
$\theta=84^o$ and $\theta=96^o$. The solid, dotted and dashed lines
correspond to Model A (LHAF), B (ADAF), and C (ADAF), respectively.}
\end{figure}
\subsection{Convective instability}

\begin{figure*}
\begin{center}
\includegraphics[width=0.8\textwidth]{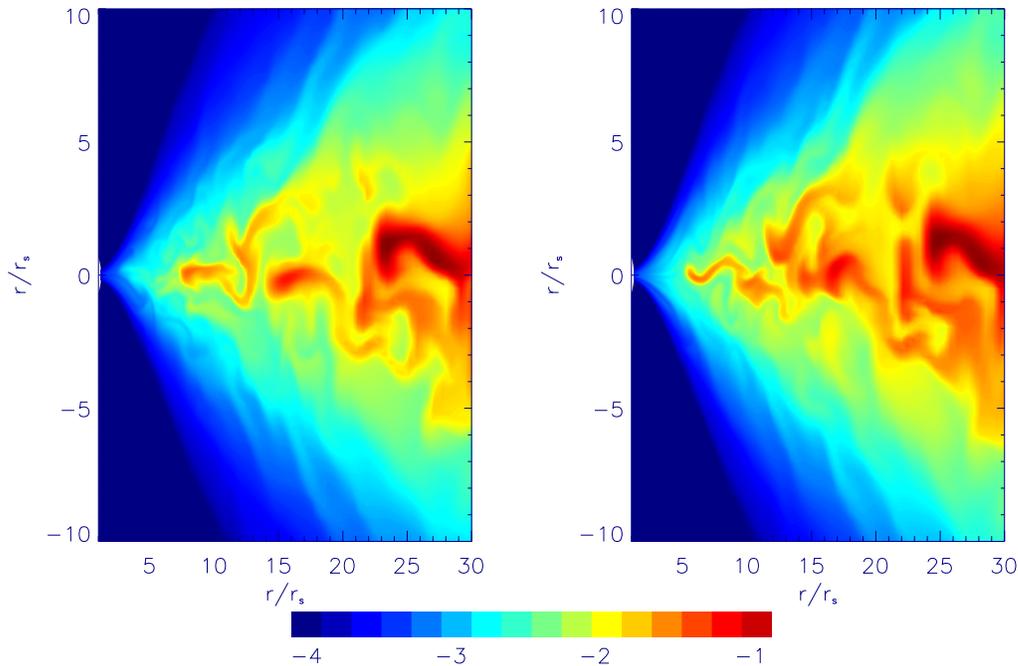}
\end{center}
\caption{Snapshots of the logarithm of the density at t=15 orbits
for Model A (left) and B (right). }
\end{figure*}

The condition for convective instability in a rotating accretion
flow is:
\begin{equation}
N^2_{\rm eff}\equiv
N^2+\kappa^2=-\frac{1}{\rho}\frac{dP}{dR}\frac{d{\rm
ln}(P^{1/\gamma}/\rho)}{dR}+\kappa^2<0,
\end{equation} where $N$ is
the usual Brunt-V\"ais\"al\"a frequency and $\kappa$ is the
epicyclic frequency which is equal to rotation angular velocity
$\Omega$ for nearly Keplerian rotation (Narayan \& Yi 1994). For a
non-rotating flow, $\kappa=0$, this condition is then equivalent to
an inward increase of entropy, which is the well-known Schwarzschild
criterion. For a rotating flow, the inward increase of entropy is a
necessary condition for convective instability. In other words, the
flow must be convectively stable if the entropy decreases inwardly.

As we state in introduction, numerical simulations show that an ADAF
without radiation is convectively unstable (IA99; SPB99;
Igumenshchev \& Abramowicz 2000). This is because of the inward
increase of entropy in an ADAF (Narayan \& Yi 1994). Following
SPB99, if we define the mass inflow, outflow, and net rates as,
\begin{equation}
 \dot{M}_{\rm in}(r) = 2\pi r^{2} \int_{0}^{\pi} \rho \min(v_{r},0)
   \sin \theta d\theta,
\end{equation}
\begin{equation}
 \dot{M}_{\rm out}(r) = 2\pi r^{2} \int_{0}^{\pi} \rho \max(v_{r},0)
    \sin \theta d\theta,
\end{equation}
\begin{equation}
\dot{M}_{\rm net}(r)=\dot{M}_{\rm in}(r)-\dot{M}_{\rm out}(r),
\end{equation}
as a result of convective instability, both the mass inflow and
outflow rates decrease inward (ref. Fig. 6 in SPB99). This is
physically because of convective outflow (note that these outflow
can't escape to infinity because of their negative value of
Bernoulli parameter) and gas circulation in convective eddies. This
is one of the most important results of previous hydrodynamical
simulations of hot accretion flows in the past decade.

\begin{figure*}
\begin{center}
\includegraphics[width=0.8\textwidth]{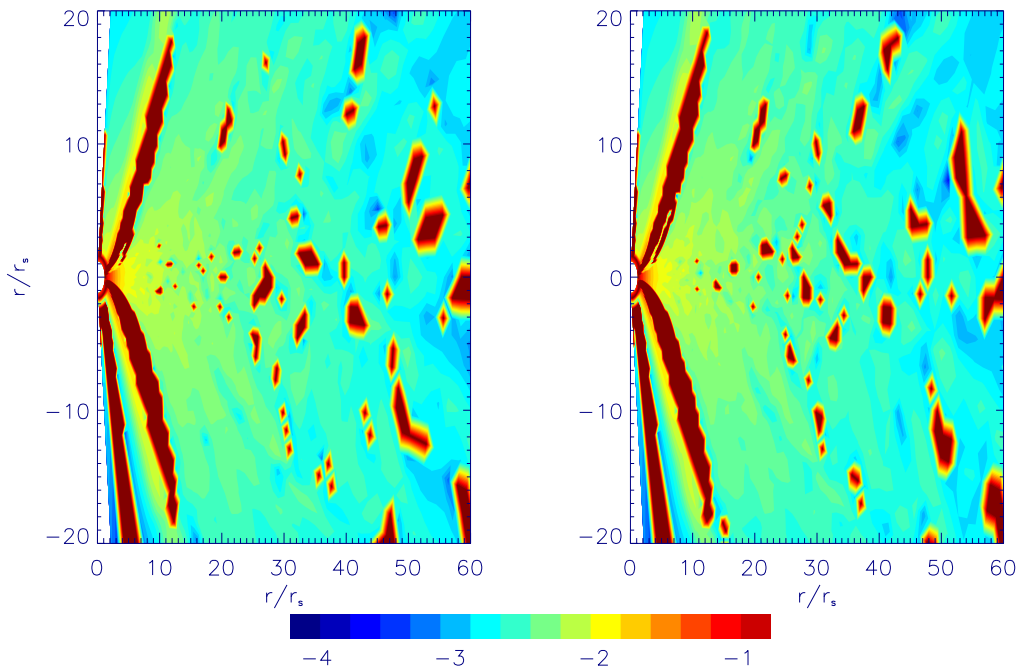}
\end{center}
\caption{Snapshots of $N_{\rm eff}^2$ at t=15 orbits for Model
A(left) and B (right). The red region denotes $N_{\rm eff}>0$, all
other colors denote $N_{\rm eff}<0$. We can see that for both
models, $N_{\rm eff}^2<0$ in most of the region. The labels of the
color bar is the logarithm of the absolute value of $N_{\rm eff}$.}
\end{figure*}

All previous studies neglect radiation. In reality, radiation is
strong in many, if not most, of black hole systems. A natural
question is then whether the flow is the convectively stable or not
in this case\footnote{We neglect the thermal instability in our
discussion, because even though an LHAF is thermally unstable, the
growth timescale of perturbation is longer than the accretion
timescale (Yuan 2003).}. When radiation is important, it is possible
that the convective instability may become weaker or even becomes
convectively stable. This is because, qualitatively, radiation plays
a similar role to convection in terms of carrying away the
dissipated energy. Quantitatively, the inclusion of radiation will
change the gradient of entropy. The energy equation (eq. 3) can be
re-written as:
\begin{equation}
\rho T \left(\frac{\partial S}{\partial t}+\mathbf v\cdot \nabla
S\right)={\mathbf T}^2/\mu-Q_{\rm rad}^- \end{equation}  So the
inclusion of radiation will make the radial profile of entropy
flatter. If radiation is strong enough so that $Q^-_{\rm rad}
> Q_{\rm vis} (\equiv {\mathbf T}^2/\mu)$, as in an LHAF, the gradient
of entropy will change sign. For a one-dimensional LHAF, this
implies \be \rho T v_r \frac{\partial S}{\partial r}={\mathbf
T}^2/\mu-Q^-_{\rm rad} < 0.\ee This means that the entropy of an
LHAF will decrease inwardly ($v_r<0$). Therefore, Yuan (2001)
speculates that different from an ADAF, an LHAF should be
convectively stable.

However, our two-dimensional numerical simulations indicate that
this is not true. We find that Model A is also convectively
unstable. The two plots in Figure 5 show the snapshots of the
logarithm of the density at $t=15$ orbits for Models A (left plot)
and B (right plot), respectively. It is hard to tell any significant
difference between them. In both plots the large-amplitude
fluctuations at small scales which is associated with convective
motion are evident. We calculate the Brunt-V\"ais\"al\"a frequency
$N_{\rm eff}$ (eq. 8) and do find $N_{\rm eff}^2 < 0$ in most of the
region of the accretion flow for both Model A and B, as shown by
Figure 6.

We also calculate the mass fluxes of the three models based on eqs.
(9-11). The results are shown in Figure 7. The black, red, and blue
lines correspond to Model A, B, and C, respectively. In each model,
the solid, dashed, and dotted lines denote the rates of inflow,
outflow, and net accretion, respectively. In all cases, the net
accretion rate is constant with radius, which indicates that our
simulations have achieved quasi-steady state. In all three models,
the inflow rate $\dot{M}_{\rm in}$ and outflow rate $\dot{M}_{\rm
out}$ decrease inward because of the convective instability, as we
expect, with almost the same ``steepness''. Quantitatively, the net
accretion rates of Models A, B, and C are $\dot{M}_{\rm
net}=\dot{M}_{\rm in}(r)-\dot{M}_{\rm out}(r)=6.4\times 10^{-4},
3.97 \times 10^{-6}$ and $4.02 \times 10^{-8}\dot{M}_{\rm Edd}$,
respectively. If the radiation does not affect the strength of
convection, we would expect that the net accretion rates differ by
exactly 100 times from Model A to B, and B to C, respectively. So
such a result indicates that radiative cooling does weaken the
convective instability. Correspondingly we find that the ratio of
the energy flux associated with convection and with advection is
moderately weaker in Model A than those of Model B and C. But we
want to emphasize that such an effect is very weak.

\begin{figure}
\begin{center}
\includegraphics[width=0.5\textwidth]{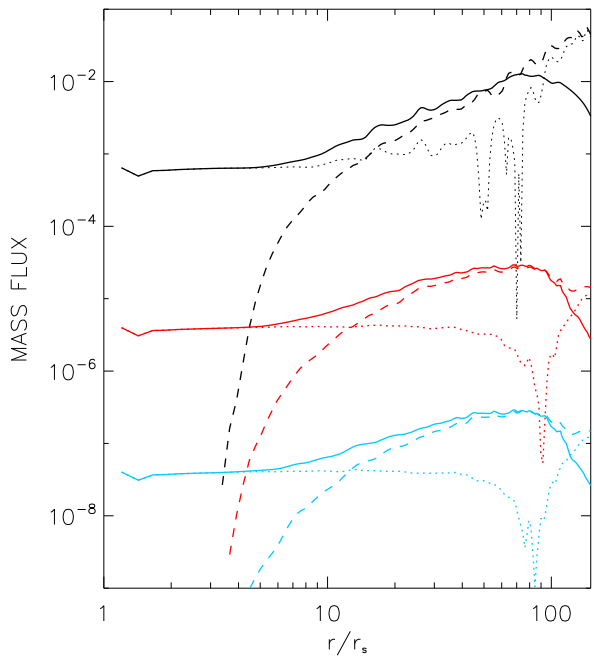}
\end{center}
\caption{Time-averaged mass fluxes measured in units of Eddington
accretion rate. The black, red, and blue lines correspond to Model
A, B, and C, respectively. In every model, the solid, dotted and
dashed lines denote the inflow rate, net accretion rate and outflow
rate, respectively.}
\end{figure}

The result that Model A is also convectively unstable is in conflict
with the one-dimensional prediction. To examine the reason of
convective instability of Model A, in Figure 8 we show the radial
and angular structures of entropy ($S={\rm ln} P-
\gamma/(\gamma-1){\rm ln}T$) of the three models. The radial
structure along the equatorial plane is shown in the upper panel
(the result along $\theta=45^{\circ}$ is similar). The middle and
lower panels show the angular structure at $r=20$ and $40 r_s$,
respectively.  The entropy increase inward in most of the region of
Model B and C, as we expect. But surprisingly we find that this is
also the case of Model A for $r<110 r_s$, although $f<0$ for
$r>30r_s$ (ref. Figure 4). {\em It is the increase of entropy along
the equatorial plane (the direction of the gravitational force) that
results in the convective instability of Model A}.

The inward increase of entropy in Model A is not in contradiction
with the negative value of advection factor $f$. For a steady
two-dimensional LHAF, the energy equation (12) reduces to,
\begin{equation}\rho T \mathbf v\cdot \nabla S \equiv \rho T \left(v_r\frac{\partial
S}{\partial r}+v_{\theta}\frac{\partial S}{\partial
\theta}\right)=Q_{\rm vis}-Q^-_{\rm rad}<0.
\end{equation} From Figure 8 we see that $\partial S/\partial r<0$, $\partial
S/\partial \theta<0$ (for $\theta<\pi/2$) and $\partial S/\partial
\theta>0$ (for $\pi>\theta>\pi/2$). Eq. (14) is satisfied because we
have $v_r<0$ and $v_{\theta}>0$ (for $\theta<\pi/2$) and
$v_{\theta}<0$ (for $\pi>\theta>\pi/2$). If without the
$v_{\theta}(\partial S/\partial \theta)$ term, i.e, for the
one-dimensional case, from $f<0$ we would have $\partial S/\partial
r >0$, i.e., the flow is convectively stable.

\begin{figure}
\begin{center}
\includegraphics[width=0.5\textwidth]{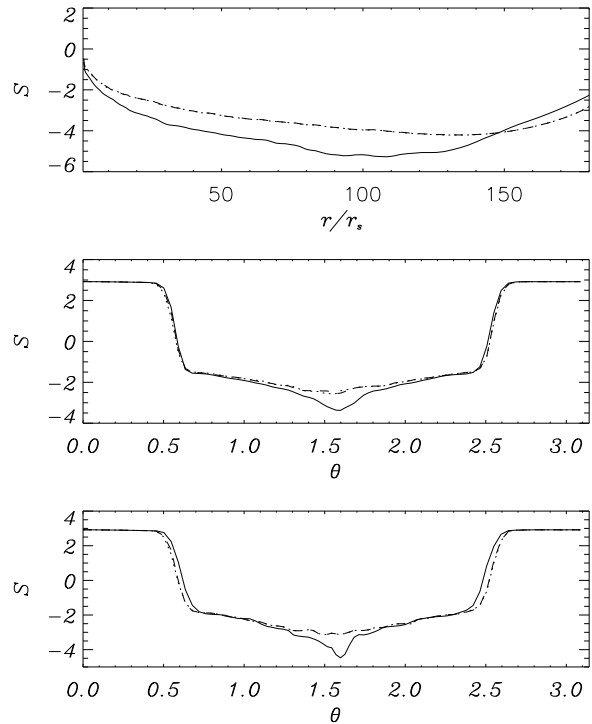}
\end{center}
\caption{The radial (upper panel) and angular (lower two panels)
structure of the specific entropy of Models A, B and C. The middle
and lower panels show the angular structure at $r=20$ and $40 r_s$,
respectively. The solid, dotted, and dashed lines show Model A, B,
and C, respectively. For the radial structure, the solution is
averaged over angle between $\theta=84^{\circ}$ and
$\theta=96^{\circ}$.}
\end{figure}

\subsection{Varying the form and amplitude of shear stress}

All the simulations described so far are based on the viscosity
stress description of $\nu\propto \rho$. In this case,
$\rho(r)\propto r^0$ (SPB99). We have also tried the description of
$\nu\propto r^{1/2}$ which is the usual ``$\alpha$'' description.
This is the description adopted in IA99 and Run K in SPB99. Under
this description, the density profile is steeper, $\rho \propto
r^{-1/2}$ (SPB99). The higher density at the innermost region makes
the bremsstrahlung radiation much stronger compared to the former
description. From eq. (14) we see that $\mathbf v\cdot\nabla S$ will
become ``more negative''. We therefore need to check the convective
instability in this case.

For this aim, we simulate three models with this kind of viscous
stress description. All other parameters are the same as Model A, B,
and C\footnote{Because of the difference of the viscous stress
description, the accretion rates of each three models here are $\sim
3$ times smaller compared to those in \S3.1, although the density of
the initial torus is the same.}. Specifically, for Model A which has
the highest accretion rate, the advection factor $f<0$ for
$r>40r_s$. Our simulation results indicate that all three models are
again convectively unstable, with $N_{\rm eff}^2<0$ in most of the
region, as in the cases presented in \S3.2. Because of the
convective instability, the mass fluxes again decrease inward, as
shown by Fig. 9. The physical reason is same to the cases of the
former viscous stress description. Figure 10 shows the radial and
angular profiles of entropy. We see from the figure that the flow
adjusts itself so that the entropy increases radially inward,
although the advection factor $f<0$.

\begin{figure}
\begin{center}
\includegraphics[width=0.5\textwidth]{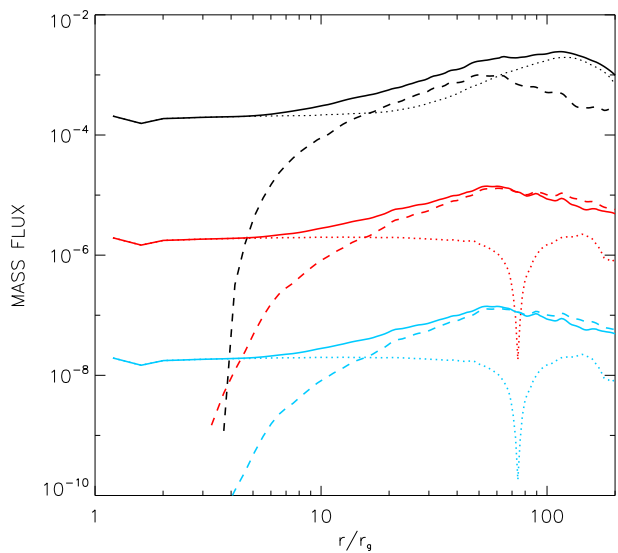}
\end{center}
\caption{Time-averaged mass fluxes measured in units of Eddington
accretion rate for viscosity stress description $\nu\propto
r^{1/2}$. The black, red, and blue lines correspond to Model A, B,
and C, respectively. In every model, the solid, dotted and dashed
lines denote the inflow rate, net accretion rate and outflow rate,
respectively.}
\end{figure}

\begin{figure}
\begin{center}
\includegraphics[width=0.5\textwidth]{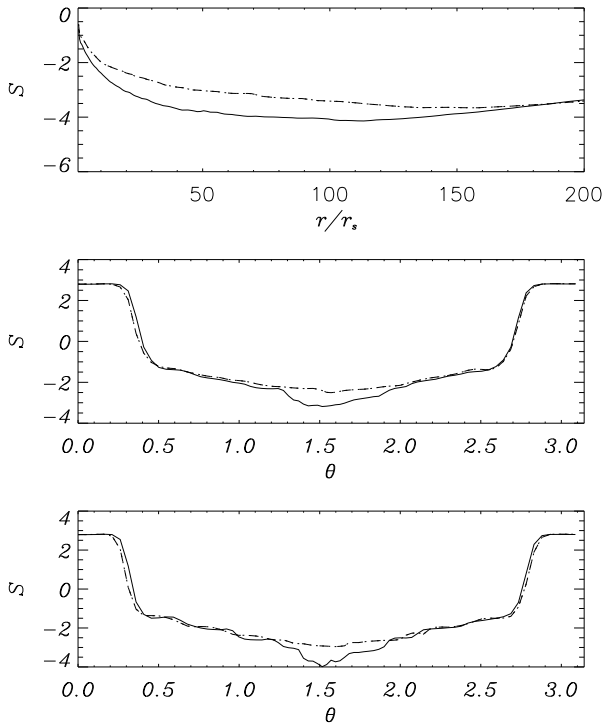}
\end{center}
\caption{The radial (upper panel) and angular (lower two panels)
structure of the specific entropy of Models A, B and C for the
$\nu\propto r^{1/2}$ viscosity stress description. The middle and
lower panels show the angular structure at $r=20$ and $40 r_s$,
respectively. The solid, dotted, and dashed lines show Model A, B,
and C, respectively. For the radial structure, the solution is
averaged over angle between $\theta=84^{\circ}$ and
$\theta=96^{\circ}$.}
\end{figure}

IA99 find that when the viscous stress is large, $\alpha\sim 1$ in
the language of the $\alpha$ description ($\nu\propto r^{1/2}$), the
convective instability disappears and the accretion flow shows
powerful unbound bipolar outflow structure. SPB99 also simulated the
case of large viscous stress. Different from IA99, they find that
convective instability still exists. They speculate that the
discrepancy may be because that the poloidal components of the shear
stress tensor adopted in IA99 suppress the convection. But the
simulation of SPB99 is based on the $\nu\propto \rho$ description
which is different from IA99. Although this is not expected to bring
significant difference on the existence of convective instability,
it is more direct to adopt the $\nu\propto r^{1/2}$ stress
description to check the result. We adopt such a viscous description
and find that in the case of a large $\alpha$, when all the
components of the shear stress are adopted, we do recover the
results of IA99, namely convective instability dies off and the
accretion flow shows bipolar outflow structure. However, when only
the azimuthal components are adopted as in SPB99, we find that the
accretion flow is still convectively unstable even though $\alpha$
is large.

\section{Summary and discussion}

Previous works without including radiation indicate that the flow is
strongly convectively unstable. We have investigated the effects of
radiation on the properties of hot accretion flows based on
two-dimensional hydrodynamical numerical simulations. Special
attention is paid to the convective instability of accretion flow
when radiative cooling is strong. This question is important since
it determines the radial profile of accretion rate and thus the
observational appearances of black hole system, and the evolution of
black hole mass and spin.

We consider the effects of radiation by including a bremsstrahlung
radiation term in the energy equation. Three models (Model A, B, and
C) are adopted with mass accretion rates spanning four orders of
magnitude. We find that with the increase of accretion rate, the
flow becomes cooler thus the mass more concentrated onto the
equatorial plane. In all cases, the Bernoulli parameter is always
negative, except in the small region close to the poles. This
results in very weak unbound outflow, with the mass flux being only
$1\%$ of the inflow rate.

Radiation is very weak in Model B and C thus they are convectively
unstable. For the model with the highest accretion rate (Model A),
the radiative cooling is so strong that it is larger than the
viscous heating in most of the region, i.e, the advection factor is
negative (ref. Figure 4). This model thus belongs to the regime of
luminous hot accretion flow (LHAF; Yuan 2001). LHAF is speculated to
be convectively stable in the one-dimensional analysis because the
entropy decreases inward (Yuan 2001). However, our simulations
indicate that such a one-dimensional result is not correct. Same
with Model B and C, Model A is also convectively unstable (Figures
5\&6). As a result, the inflow and outflow rates are decreasing
inward, with almost the same ``steepness'' of the profile of mass
flux for all three models (Figure 7). We find that the reason is
because the entropy increases radially inward in two-dimensional
case (Figure 8).

We only consider bremsstrahlung radiation in our work. As a result,
the advection factor is always positive in the most interesting
innermost region, even though a large accretion rate is adopted. It
will be interesting to investigate the convective instability of a
model with a negative $f$ in the whole region of the accretion flow.
This can be achieved in principle by adopting a higher accretion
rate. But the problem is that if a much larger accretion rate were
adopted, the radiation at the outer region will become so strong
that the hot accretion flow will collapse. One way to avoid this
problem is to include synchrotron radiation and especially its
Comptonization. Since they are much stronger than the bremsstrahlung
radiation at the innermost region of the flow, the advection factor
at small radii will easily become negative, even smaller than that
in the outer region (ref. Yuan 2001). The convective stability in
this case will be an important project.

In this paper we consider the effects of radiation only by including
a radiative cooling term in the energy equation. Further improvement
on radiation can be done by calculating the radiative transfer. For
hot accretion flows, the radiation pressure is always smaller than
the gas pressure thus the effect of radiation in the momentum
equation can be neglected. But the interaction of energy between
radiation and gas will be important when the accretion rate is
large. Since the radial optical depth of accretion flow is less than
unity, the radiation produced at one radius can propagate for a
large distance and heat or cool gas at other radius via Compton
scattering. As a result, the dynamics of the accretion flow will be
significantly changed (e.g., Ostriker et al. 1976; Cowie et al.
1978; Park \& Ostriker 1999; 2001; 2007; Yuan, Xie \& Ostriker 2009;
Xie et al. 2010). At the outer region, such Compton scattering plays
a heating role. It is found that the self-consistent solution
including this effect can only exist below a certain accretion rate,
because the flow beyond a certain large radius will be heated above
the virial temperature thus the accretion is suppressed\footnote{Any
significant energy flux from the inner to the outer region of an
accretion flow can reduce the mass inflow rate. In addition to
convection which is the focus of this paper, radiation is another
mechanism of transferring energy outward (Ostriker et al. 1976).
Other examples include viscous stresses (Blandford \& Begelman 1999)
and thermal conduction (Tanaka \& Menou 2006; Johnson \& Quataert
2007).}. Above this value, the black hole will oscillate between an
active and inactive phase (Cowie et al. 1978; Yuan, Xie \& Ostriker
2009). Time-dependent numerical simulation is required to check
these analytical results. Regarding the convective instability,
Compton scattering will heat gas at large radii but cool gas at
inner region, thus the entropy of the flow will become larger at
outer region but smaller at inner region. So the entropy will
decrease faster inward compared to Model A in the present paper.
Whether this is sufficient to change the sign of the gradient of
entropy and thus stabilize the accretion flow against the convective
instability is another interesting project.

We adopt an anomalous shear stress to transfer the angular momentum.
In reality, magnetic field must exist and the turbulence associated
with the magnetorotational instability (MRI) is believed to be the
origin of viscosity. A question is then whether the hydrodynamical
analysis adopted in the present paper is applicable to a
magneto-hydrodynamical (MHD) flow. There have been some discussions
in the literature on this point (Hawley, Balbus, \& Stone 2001;
Balbus \& Hawley 2002; Narayan et al. 2002). Narayan et al. (2002)
show that if the magnetic field saturates at a value sufficiently
below equipartition, convective fluctuations can be applicable to an
MHD accretion flow (since long-wavelength convective fluctuations
can fit inside the accretion disk). Unfortunately, the saturated
magnetic field in MRI simulation of accretion flow seems to be not
universal, e.g., depending on the initial configuration of the
magnetic field (Stone \& Pringle 2001; Machida et al. 2001).
Although our focus of the present work is on the main body of
accretion flow, we would like to mention in this context that the
coronal region of accretion flow is known to be strongly magnetized,
with magnetic field there exceeding the equipartition strength
(e.g., Miller \& Stone 2000; Beckwith, Hawley \& Krolik 2008). So
our hydrodynamical analysis does not apply in that region.

We mention in the introduction that the luminosity of black hole
X-ray binary almost remains constant before and after the state
transition. If the hot accretion flow is still convectively unstable
after all the above-mentioned effects have been taken into account,
which implies that the accretion rate decreases inward from the
transition radius $R_{\rm tr}$, the value of $R_{\rm tr}$ must be
small so that the accretion rates of the inner hot and outer cool
accretion flows are similar. This puts an independent constraint on
the mechanism of transition from an outer cold disk to an inner hot
accretion flow.

\section{ACKNOWLEDGMENTS}

We thank Woong-Tae Kim and Ramesh Narayan on their valuable comments
on the paper. We also benefit significantly from the advices of
Woong-Tae Kim on the ZEUS code. The idea of considering radiative
cooling is stimulated from the discussion with James Stone. We also
thank the referee for his/her constructive suggestions which greatly
improve the presentation of the paper. This work was supported in
part by the Natural Science Foundation of China (grants 10773024,
10833002, 10821302, and 10825314), the National Basic Research
Program of China (973 Program 2009CB824800), and the CAS/SAFEA
International Partnership Program for Creative Research Teams.

\end{document}